# The impact of niobium doping upon the magnetotransport properties of the oxygen-deficient perovskite SrCo$_{1-x}$Nb$_x$O$_{3-\delta}$

T. Motohashi,* V. Caignaert, V. Pralong, M. Hervieu, A. Maignan, and B. Raveau

Laboratoire CRISMAT, UMR CNRS ENSICAEN 6508, 6 bd Maréchal Juin
14050 CAEN Cedox 4 France

The oxygen-deficient perovskite cobaltite SrCo$_{1-x}$Nb$_x$O$_{3-\delta}$ was synthesized by direct solid-state reaction and its magnetotransport properties were investigated. This cobaltite exhibits an unusual ferromagnetic behavior with a transition temperature $T_m$ = 130-150 K and a spin glass like behavior below $T_m$. Importantly, this phase reaches a large magnetoresistance (MR) value, MR $\equiv$ -($\rho_H$ - $\rho_0$) / $\rho_0$ = 30% at 5 K in 7 T. The large MR effect is believed to be related to the disordered magnetic state induced by the Nb-for-Co substitution.

* Corresponding author
Permanent address: Materials and Structures Laboratory, Tokyo Institute of Technology, 4259 Nagatsuta, Midori-ku, Yokohama 226-8503, Japan.
E-mail: t-mot@msl.titech.ac.jp
Phone: +81-45-924-5318
Fax: +81-45-924-5339

One of the most attractive phenomena in functional transition-metal oxides deals with their large magnetoresistance (MR) properties. The first example is the colossal MR effect in perovskite manganites, for which the MR magnitude, defined as MR ≡ $-(\rho_H - \rho_0)/\rho_0$, often surpasses 90% (for a review see Ref. 1). Besides, cobaltites with the perovskite structure were also found to exhibit large MR values, as illustrated for $La_{1-x}Sr_xCoO_3$ [2-4] and the ordered oxygen-deficient perovskite $Ln BaCo_2O_{5.4}$ [5], whose MR ranges from 12.5% to 50% in 7 T at low temperatures.

In contrast to perovskite manganites, the magnetotransport properties of cobaltite perovskites do not seem to cover a broad range of composition. For instance, no MR effect has been reported to date about the stoichiometric $SrCoO_3$ synthesized using electrochemical oxidation [6] or high pressure technique [7] and the MR effect was found to be negligible in the oxygen-deficient perovskite $SrCoO_{2.75}$, prepared by a two-step method [8]. However, a MR value of 12% in 7 T at 5 K was recently reported for the Sc-substituted oxygen-deficient perovskite $SrCo_{1-x}Sc_xO_{3-\delta}$ [9]. This suggests that the introduction of a $d^0$ cation, like $Sc^{3+}$, on the cobalt sites, not only stabilizes the perovskite structure but also is susceptible to induce large MR. We have thus explored the possibility to stabilize the cobalt perovskite by a $d^0$ cation with a higher valency such as $Nb^{5+}$. In this letter, we report on the magnetotransport properties of the oxygen-deficient perovskites $SrCo_{1-x}Nb_xO_{3-\delta}$. We show that the latter exhibits resistivity, one order of magnitude smaller than for the Sc-substituted cobaltite and a MR value three times higher. We also suggested that this high MR value is closely related to the presence of a disordered magnetic state induced by the Nb-for-Co substitution.

Samples of $SrCo_{1-x}Nb_xO_{3-\delta}$ with $x = 0.05, 0.10, 0.15$, and $0.20$ were prepared from powder mixtures of $SrCO_3$, $Co_3O_4$, and $Nb_2O_5$ with appropriate ratios which were calcined in flowing $O_2$ at 800°C for 12 h. This calcined powder was ground, pelletized, and fired in flowing $O_2$ at 1200°C for 12 h, followed by slow cooling down to room temperature. A part of the samples was subsequently post-annealed at 500°C for 24 h in high-pressure oxygen (10 MPa). X-ray powder diffraction (XRPD) analysis (detailed elsewhere [10]) indicated that all the samples were single phase with the perovskite structure, except for the post-annealed $x = 0.05$ sample which was partially decomposed after annealing, leading to the formation of $Sr_6Co_5O_{15}$ as an impurity. The lattice parameter of the cubic subcell was found to linearly increase with $x$ from 3.8560 Å for $x = 0.05$ to 3.8797 Å for $x = 0.20$ in the as-synthesized samples and from 3.8648 Å for $x = 0.10$ to 3.8776 Å for $x = 0.20$ in the $O_2$-annealed samples.

The electron diffraction (ED) analysis confirmed a good crystallinity of the compounds. The ED patterns exhibit a set of intense reflections characteristic of a perovskite-related structure, the presence of weaker extra spots suggesting additional ordering phenomena which will be reported elsewhere. The energy dispersive analysis (EDS) performed with a kevex analyzer mounted on a transmission electron microscope, JEOL 200 CX, confirmed the nominal cationic compositions and the homogeneity of the samples. The chemical analysis using redox titration leads to oxygen contents ranging from $O_{2.74}$ ($x = 0.05$, as-synthesized) to $O_{2.83}$ ($x = 0.20$, $O_2$-annealed). It was shown that the cobalt valency decreases from +3.39 for $x = 0.05$ to +3.30 for $x = 0.20$ in the as-synthesized samples and slightly increases by $O_2$-annealing, e.g. +3.33 for $x = 0.20$ [10].

The temperature dependence of resistivity $\rho$ (Fig. 1), measured using a four-point-probe apparatus (PPMS; Quantum Design) shows a similar behavior, to that observed for SrCo$_{1-x}$Sc$_x$O$_{3-\delta}$ [9], i.e. $\rho$ increases systematically with the Nb content $x$, due to the decrease in the number of conducting paths, interrupted by Nb$^{5+}$ species. In both cases, $\rho$ decreases by O$_2$ annealing, in agreement with the increase in Co$^{4+}$ content. Nevertheless, it must be emphasized that for low $x$ values ($x \leq 0.10$) the resistivity of the Nb-samples at low $T$ is an order of magnitude smaller than that of the Sc-phase.

Magnetic properties were investigated using a vibrating sample magnetometer and a dc-ac SQUID magnetometer. The dc magnetization ($M$) curves show a significant increase of $M$ at low $T$ (Fig. 2), indicating the presence of ferromagnetic interactions. The magnetic transition temperature $T_m$ = 130-150 K, determined by means of ac-susceptibility measurements (a representative result is presented in Fig. 3), does not strongly depend on the Nb content ($x$) while it slightly increases by O$_2$ annealing. The absolute value of $M$ rapidly decreases with increasing $x$, being consistent with the fact that non-magnetic Nb$^{5+}$ cations dilute the magnetic interaction. The magnetic behavior gets enhanced by O$_2$ annealing for all the $x$ values, due to the increase in the concentration of magnetic Co$^{4+}$ species.

These magnetic data show that SrCo$_{1-x}$Nb$_x$O$_{3-\delta}$ is, like the Sc-substituted cobaltites, clearly different from a typical ferromagnet. Its $M$ value at 5 K is not indeed saturated in 5 T, with a value of ~0.8 $\mu_B$/Co, i.e. much smaller than that reported for SrCoO$_3$, 1.8-2.1 $\mu_B$/Co [6,7]. Nevertheless, it is worth pointing out that this value is higher than for the Sc-phase, 0.3 $\mu_B$/Co [9]. In fact the magnetic susceptibility in the field-cooled process (inset of Fig. 2) and the frequency dependent ac-susceptibility peak (Fig. 3), clearly indicate a disordered magnetic state, characteristic of a spin glass or a cluster glass.

The $\rho(T)$ curves in 7 T (Fig. 1) show large MR effects. The MR vs $H$ measurements were performed at several temperatures. A typical result is shown in Fig. 4. The MR value monotonically increases with decreasing temperature. A large hysteresis is seen only at 5 K for all the samples. The MR effect closely correlates with the $M(H)$ curve (inset of Fig. 4): the resistivity is highest at the coercive field, $\approx$0.6 T, suggesting a maximum number of magnetic domain boundaries as shown for a $Sr_2CoO_4$ thin film [11]. The MR values (in 7 T) at 5, 50, 100, and 150 K were plotted in Fig. 5 against the Nb content, $x$. The MR value in 7 T is as large as 30% at 5 K and still 27% even at 50 K for the $x = 0.15$ and 0.20 $O_2$-annealed samples. This value is three times larger than that reported for $SrCo_{1-x}Sc_xO_{3-\delta}$ [9].

The monotonic increase in MR below $T_m$ implies that the negative MR effect mainly stems from spin-dependent scattering of carriers which is related to the ferromagnetic state. The correlation between MR($H$) and $M(H)$ also supports this hypothesis. Moreover, for $x = 0.2$ the magnetization is significantly enhanced by $O_2$ annealing, and accordingly, the MR magnitude increases (Figs. 2 and 5). Thus, it is obvious that there exists a threshold in the ferromagnetic interaction for the appearance of a large MR effect. However, ferromagnetism is not the only factor required for large MR as shown by comparing with the $SrFe_{1-x}Co_xO_{3-\delta}$ series [8] which exhibits smaller MR effect (~12%) in spite of large magnetic moment (> 1 $\mu_B$/f.u.). More probably, the magnetic disorder induced by Nb-for-Co substitution, leading to a spin glass behavior, renders an additional contribution in the enhanced MR: i.e. the field-induced ferromagnetism further reduces spin-dependent scattering of carriers. The comparison of the $x = 0.10$ as-synthesized and $x = 0.15$ annealed samples (or the 0.15 as-synthesized and 0.20 annealed samples), which both have almost the same $\rho$ and $M$ values, but exhibit different MR values of 26% and 30% (24% and 31%) at 5 K, respectively, is in agreement with this view point.

Finally, it must be emphasized that this Sr-rich series $SrCo_{1-x}Nb_xO_{3-\delta}$ and La-rich series $La_{1-x}Sr_xCoO_3$ exhibit very different magnetic and MR properties. For instance, $La_{0.7}Sr_{0.3}CoO_3$ is a ferromagnetic metal (Curie temperature $T_C$ = 220-240 K) with only small MR values ($\approx$ 9% in 6 T) [3,4] whereas $SrCo_{0.8}Nb_{0.2}O_{2.83}$ has a disordered magnetic state with much larger MR, in spite of the same cobalt valency of these two oxides (= +3.3). Clearly, both magnetic dilution through Nb-for-Co substitution and oxygen vacancies are important ingredients to induce large MR in the perovskite cobaltites.

In summary, we have shown the possibility to synthesize Nb-substituted oxygen-deficient perovskites $SrCo_{1-x}Nb_xO_{3-\delta}$, with large negative MR, reaching 30% at 5 K in 7 T. The excellent MR property of this unusual weak ferromagnet with $T_m$ = 130-150 K may originate from the existence of the disordered magnetic state, induced by Nb-for-Co substitution. Detailed investigations of magnetism and structure of the present compound will be necessary to further understand these phenomena.

Figure captions

Fig. 1:

Temperature dependence of resistivity ($\rho$) for the as-synthesized (a) and $O_2$-annealed (b) $SrCo_{1-x}Nb_xO_{3-\delta}$ samples. The solid and broken curves show $\rho$ data recorded in 0 T and 7 T, respectively.

Fig. 2:

Temperature dependence of magnetization (*M*) for the $SrCo_{1-x}Nb_xO_{3-\delta}$ samples. The data were recorded in the zero-field-cooling mode (0.25 T) using a vibrating sample magnetometer. The inset shows susceptibility data for the $x = 0.10$ $O_2$-annealed sample. The data were recorded in 0.3 T using a SQUID magnetometer.

Fig. 3:

Temperature dependence of ac-susceptibility, (a) $\chi'$ and (b) $\chi''$, for the $O_2$-annealed $x = 0.10$ sample. The data were recorded in an ac-field of 3 Oe with frequencies of 1, 10, and 100 Hz. The transition temperature $T_m$ is marked with an arrow.

Fig. 4:

MR vs *H* plots for the $x = 0.15$ $O_2$-annealed sample at 5, 50, 100, 150, and 200 K. In each measurement, the sample was zero-field-cooled, and the magnetic field was swept from 0 to 7 T, then 7 to -7 T. A large hysteresis is seen only at 5 K. The inset shows the *M* vs *H* loop at 5 K for the same sample.

Fig. 5:

The MR values (in 7 T) at 5, 50, 100, and 150 K against the Nb content, $x$. The MR values were obtained from the MR vs $H$ measurements. Open and closed symbols denote the as-synthesized and $O_2$-annealed samples, respectively.

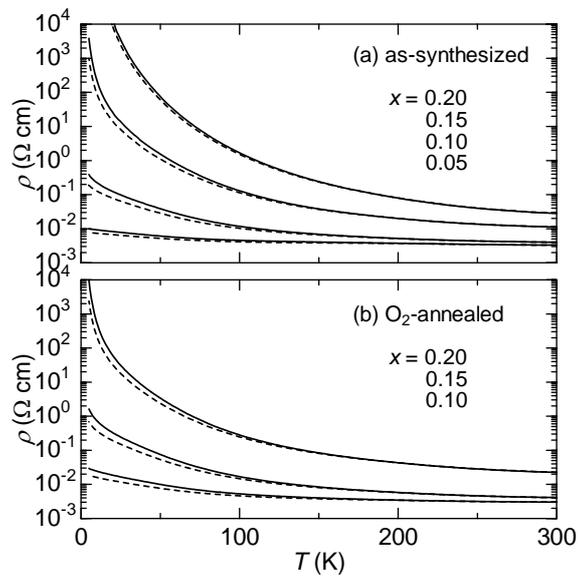

Fig. 1: Motohashi *et al.*

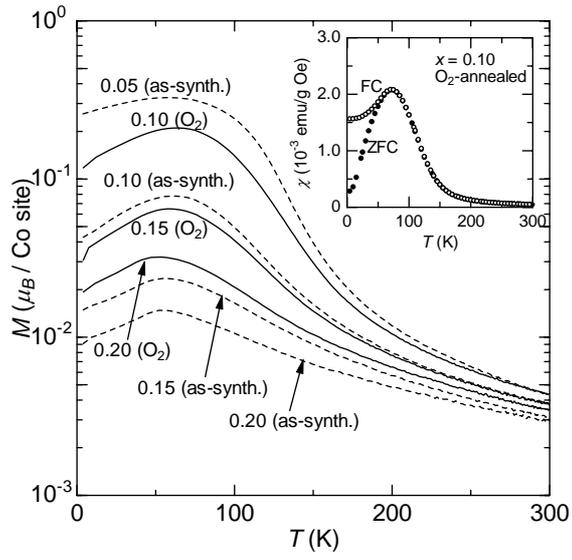

Fig. 2: Motohashi *et al.*

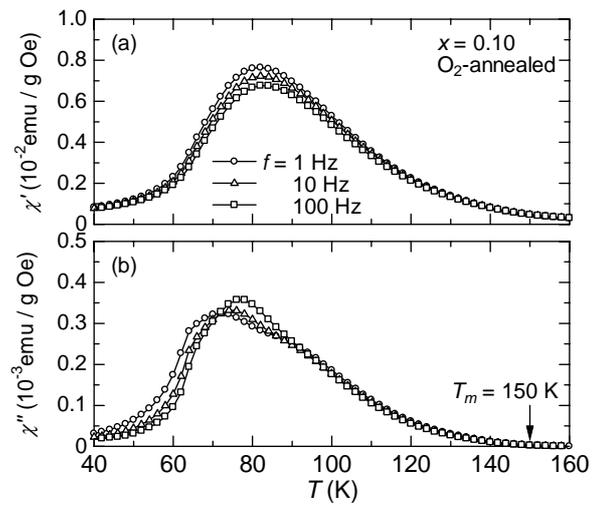

Fig. 3: Motohashi *et al.*

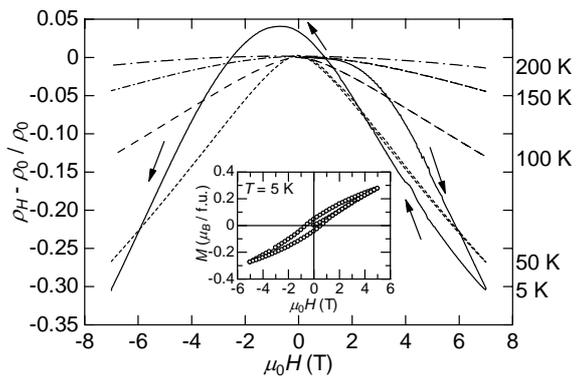

Fig. 4: Motohashi *et al.*

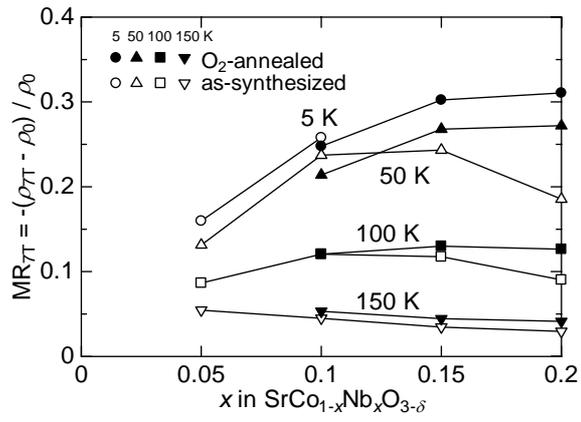

Fig. 5: Motohashi *et al.*